\documentclass[a4paper,12pt]{article}
\usepackage[utf8]{inputenc}
\usepackage[T2A]{fontenc}
\usepackage[english]{babel}
\usepackage{geometry}
\usepackage{fancyhdr}
\usepackage{amsmath,amsfonts,amsthm,amssymb}
\usepackage{graphicx}
\usepackage{hyperref}
\usepackage{amssymb,graphicx,float,amsmath,amsfonts,amsthm,cases}
\usepackage{float}
\usepackage{subfig}
\usepackage{mathtools}
\usepackage[symbol*]{footmisc}
\usepackage{tensor}
\usepackage[numbers]{natbib}

\begin{document}
\vspace{10pt}

\begin{center}
	{\LARGE \bf A pseudo-spectral approach to constructing rotating boson star spacetimes}
	
	\vspace{20pt}
	
	N. Sukhov$^{a,b}$\footnote[1]{\textbf{e-mail:} nsukhov@princeton.edu},
	\vspace{15pt}
	
	$^a$\textit{Department of Physics, Princeton University, Princeton, New Jersey, 08544, USA}\\
	$^b$\textit{Princeton Gravity Initiative, Princeton University, Princeton, New Jersey, 08544, USA}\\	
	\vspace{5pt}
	
\end{center}

\vspace{5pt}

\begin{abstract}
	We present a novel numerical solver for the systems of coupled non-linear elliptical differential equations. The solver partitions the computational domain into a set of rectangular pseudo-spectral collocation subdomains and is especially well-suited for working with stiff solutions such as almost shell-like solitonic boson stars. The method can be used in any number of dimensions although is practical for one- to three-dimensional problems. We apply the method to rotating and spherically symmetric solitonic boson stars and demonstrate that it displays exponential convergence. In the spherical symmetric case we explore families of almost shell-like solitonic boson stars and get the results that conform with the well-known analytic approximation \citep{Friedberg:1986tq}.
\end{abstract}

\section{Introduction}
Boson Stars (BS) represent stationary localized solutions of a system of Einstein-Klein-Gordon equations, describing a complex scalar field configuration in a curved spacetime. The BS models differ by the scalar field potential that describes the self-interaction of the scalar field and can be divided into three groups depending on the mechanism that stabilizes a star and prevents it from gravitational collapse. The first kind of stars is stabilized by a dispersion mechanism similar to Heisenberg uncertainty principle \citep{Liebling:2012fv}, the simplest kind of model belonging to this class is a mini-boson star \citep{Kaup:1968zz}. The second kind is stabilized by repulsive self-interactions between bosons that constitute the star and is represented by massive boson stars \citep{Colpi:1986ye}. The third kind of stars \citep{Friedberg:1986tq, Kleihaus:2005me} have shell-like structure with a wall that separates two regions with two different vacua. They are prevented from gravitational collapse by the surface tension in the shell \citep{Boskovic:2021nfs}. These stars are typically called Solitonic Boson Stars (SBS) due to their connection to so-called non-topological solitons or Q-balls proposed by \citep{Friedberg:1976, Coleman:1985ki}. The simplest kind of potential that can generate such stars is the sextic potential
\begin{equation}
	\label{Description:Solitonic_potential}
	V(|\Phi|^2) = \mu^2 |\Phi|^2 \left(1 - \frac{2|\Phi|^2}{\sigma_0^2}\right)^2.
\end{equation}

One of the key properties of the SBS is that by tuning the scalar field potential parameter $\sigma_0$ one can make the scalar field profile of the star very stiff making the wall between the vacua thin compared to the size of the star. There exists a well-developed analytic approximation for spherically symmetric non-rotating stars in such regime \citep{Friedberg:1986tq} that allows one to take a limit when the width of the shell becomes infinitely thin. This approximation, however, breaks down in the rotating case because the shape of the shell becomes more complicated. Finding solutions in such an extreme regime poses a numerical challenge and the rotating BS with sharp profiles haven't been thoroughly investigated. In order to be able to study such solutions one has to develop a multidomain algorithm to be able to provide adequate numerical resolution into the regions where the solution is stiff and experiences sharp transitions.

In this paper we describe a multidomain Chebyshev collocation pseudo-spectral method we have developed and implemented to solve the elliptic differential equations for such solutions with stiff shell-like features. Spectral methods are employed in a number of numerical relativity codes like SpEC \citep{SpecSXS}, COLSYS \citep{Ascher:1979iha}, KADATH \citep{Grandclement:2009ju}. The main advantage of spectral methods  over finite difference and finite volume methods for solving differential equations is their fast convergence if the solution to an equation is sufficiently smooth, as such they require many fewer grid points to reach the same level of accuracy. This comes at a cost of more complicated and non-local derivative operators and strict requirements for solution smoothness. Another advantage of Chebyshev collocation methods in particular is their intricate connection to Fourier series that allows using extremely efficient Fast Fourier Transformation (FFT) methods for certain computations.

The general construction of the method is similar to the one proposed in \citep{Pfeiffer:2002wt}; the computational domain is partitioned into rectangular subdomains and the satisfying of PDE's inside the domains and on the inter-domain boundaries is done simultaneously. However, we suggest a more general way to implement interdomain matching that allows for an absolutely generic partitioning into rectangular domains in any number of dimensions. We have implemented a version of the method that takes the number of dimensions as a parameter and can in principle work in any number of dimensions, however, since the number of domain points scale very rapidly in higher dimensions we have only explicitly tested the code for $d = 1,2,3,4$. We also use a special way of setting boundary conditions in spherical coordinates \citep{Fornberg:1995,Fornberg:1996,FornbergMerrill:1997,Trefethen:2000} to avoid getting close to coordinate singularities at the origin and the axis of symmetry by effectively extending the functions in an odd or even way across the coordinate singularity. The method is designed with the maximal possible simplicity in mind ideally to just let the user to plug in the equations and get the solutions right out.

The paper is structured as following. In the next section we recall the necessary essentials about the spectral methods and illustrate the technique used for avoiding the coordinate singularities in spherical coordinates. In Section 3 we describe the method in detail and apply it to rotating BS to confirm the convergence of the method. In Section 4 we apply the method to a 1D case of spherically symmetric BS to demonstrate the performance of the method for extremely stiff solutions as well as to introduce a few useful techniques for exploring families of BS. We conclude in section 5.

\section{Chebyshev collocation pseudo-spectral method}
We are going to give a brief review of the Chebyshev collocation method. The method relies on approximating functions with Chebyshev polynomials
\begin{equation}
	\label{Num:Chebyshev_expansion}
	f(x) = \frac{1}{2} c_0 + \sum\limits_{k = 1}^{N-1} c_k T_k (x) + \frac{1}{2}c_N T_N(x),
\end{equation}
where the polynomials are given by the formula
\begin{equation}
	T_n(x) = \cos(n \arccos x).
\end{equation}
Chebyshev polynomials are defined on the domain $x \in [-1, 1]$, to use the expansion \eqref{Num:Chebyshev_expansion} on a domain with arbitrary boundaries $y \in [a, b]$ we will use a linear map
\begin{equation}
	x = \frac{a + b - 2y}{b - a},
\end{equation}
without loss of generality we assume that $x \in [-1, 1]$ unless specified otherwise. The expression \eqref{Num:Chebyshev_expansion} is slightly different from the one usually encountered in the literature \citep{PressTeukolsky:2007}, we have added an extra $c_N$ term to illustrate better the connection with discrete Fourier transformation. A Chebyshev polynomial $T_N(x)$ of order $N$ has $N+1$ extrema located at
\begin{equation}
	\label{Num:CGL_grid}
	x_k = \cos\left(\frac{\pi k}{N}\right),\qquad k = 0, 1, \dots, N.
\end{equation}
The points \eqref{Num:CGL_grid} are called \textit{Chebyshev-Gauss-Lobatto (CGL) points}, a grid of CGL point is called \textit{Chebyshev-Gauss-Lobatto grid}.

If we evaluate the expansion \eqref{Num:Chebyshev_expansion} at the extrema $x_i$ we arrive at the formula for DCT-I discrete cosine Fourier transformation \citep{PressTeukolsky:2007}
\begin{equation}
	f(x_i) =  \frac{1}{2} \left(c_0 + (-1)^{i} c_N\right)+ \sum\limits_{k = 1}^{N-1} c_k \cos \left(\frac{\pi i k}{N}\right).
\end{equation}
DCT-I is self-inverse and thus we can evaluate the expansion coefficients $c_k$ very efficiently knowing the values of approximated function at the extrema points 
\begin{equation}
	\label{Num:DCT_coeff}
	c_k = \frac{2}{N}\left[\frac{1}{2} \left(f_0 + (-1)^{k} f_N\right) + \sum\limits_{j = 1}^{N-1} f_j \cos \left(\frac{\pi j k}{N}\right)\right].
\end{equation}
where $f_j = f(x_j)$. This connection is not accidental, indeed if we make a substitution $u = \arccos x$ the expression \eqref{Num:Chebyshev_expansion} becomes a Fourier series truncated at order $N$. The methods that represent functions with functional values at CGL gridpoints are called \textit{pseudo-spectral methods} or \textit{collocation methods}. From this point onward we are going to focus on pseudo-spectral methods exclusively.

We can reexpress the Chebyshev polynomial expansion \eqref{Num:Chebyshev_expansion} directly in terms of functional values $f_i$ on CGL grid using the DCT expression for the Chebyshev coefficients \eqref{Num:DCT_coeff} and  Dirichlet kernel formula
\footnote[1]{
	The Diriclet kernel formula is widely used in Fourier series theory and is given by
	\begin{equation}
		\frac{1}{2} + \sum\limits_{k = 1}^{n-1} \cos(k x) = \frac{\sin((n - 1/2)x)}{\sin(x/2)}.
	\end{equation}
}
\begin{equation}
	\label{Num:resum_expansion}
	f(x) = \sum\limits_{l = 0}^{N} f_l (D_N)_l (x),
\end{equation}
where the kernels $(D_N)_l(x)$ are given by
\begin{equation}
	(D_N)_l(x) = \left\{\begin{array}{lll}
		\frac{\cot(u/2) \sin(n u)}{2 N},\qquad l = 0,\\
		\frac{(-1)^{N + 1} \sin(N u) \tan\frac{u}{2}}{2 N},\qquad l = N,\\
		\begin{split}
			\frac{1}{2 N} \bigg[\frac{\sin\left[(N - 1/2) (u - \pi l/N)\right]}{\sin[1/2 (u - \pi l/N)]} +
			\frac{\sin[(N - 1/2) (u + \pi l/N)]}{\sin[1/2 (u + \pi l/N)]} + \\
			2 \cos(\pi l) \cos(l u)\bigg],\qquad 0 < l < N,
		\end{split}
	\end{array}\right.
\end{equation}
where $u = \arccos x$.

In order to construct a differential equations solver based on pseudo-spectral collocation method we need to be able to use the method to find function derivatives first. The expression \eqref{Num:resum_expansion} assures that taking any derivative of any order at any point inside the collocation domain is a linear operation in terms of the functional values $f_j$ on CGL grid, however, it will be especially useful to obtain the derivative values on CGL grid itself. Naturally, as a linear operation it can be represented as matrix multiplication. We define Chebyshev differentiation matrices as
\begin{equation}
	\label{Num:diff_formula}
	f^{(n)}_j = \sum\limits_{l = 0}^N (D^{(n)}_N)_{jl} f_l,
\end{equation}
where $f^{(n)}_l = f^{(n)}(x_l)$ is $n$-th derivative of a function $f$ at the CGL point $x_l$. Since Einstein equations are second order equations we will only need the first and the second derivative differentiation matrices. The formula \eqref{Num:resum_expansion} allows us to find explicit expressions for those matrices, it can also be found in
\citep{Canuto2010SpectralMF}. The first derivative differentiation matrix is given by
\begin{equation}
	\label{Num:diff_matrix_first}
	(D^{(1)}_N)_{jl} = \left\{\begin{array}{ll}
		\frac{c_j}{c_l} \frac{(-1)^{j+l}}{x_j - x_l}, &\qquad j \neq l,\\
		- \frac{x_l}{2 (1 - x_l^2)}, &\qquad 1 \leq i = j \leq N-1,\\
		\frac{2N^2 + 1}{6}, &\qquad i = j = 0,\\
		-\frac{2N^2 + 1}{6}, &\qquad i = j = N,
	\end{array}\right.
\end{equation}
and the second derivative matrix is given by
\begin{equation}
	\label{Num:diff_matrix_second}
	(D^{(2)}_N)_{jl} = \left\{\begin{array}{ll}
		\frac{(-1)^{j+l}}{c_l} \frac{x_j^2 + x_j x_l - 2}{(1-x_j^2)(x_j - x_l)^2}, & 1 \leq j \leq N-1,\;0 \leq l \leq N,\;j \neq l,\\
		- \frac{(N^2 - 1)(1 - x_j^2)+3}{3 (1 - x_j^2)^2}, & 1 \leq j = l \leq N-1,\\
		\frac{2}{3}\frac{(-1)^l}{c_l}\frac{(2N^2 + 1)(1 - x_l) - 6}{(1 - x_l)^2}, &j = 0,\; 1 \leq l \leq N,\\
		\frac{2}{3}\frac{(-1)^{l+N}}{c_l}\frac{(2N^2 + 1)(1 + x_l) - 6}{(1 + x_l)^2}, & j = N,\; 0 \leq l \leq N-1\\
		\frac{N^4 - 1}{15}, & j = l = 0,\; j = l = N,
	\end{array}\right. 
\end{equation}
where $x_j$ are CGL points given by \eqref{Num:CGL_grid}.

We can use the connection between Fourier series and Chebyshev polynomial expansions to apply Fourier series theory directly to Chebyshev methods \citep{Trefethen:2000} and use the following results
\begin{enumerate}
	\item If $f$ has $p - 1$ continuous derivatives in $L^2(\mathbb{R})$ for some $p \geq 0$ and a $p$-th derivative of bounded variation, then $c_k = O(|k|^{-p-1})$
	as $|k| \to \infty$:
	\item If $f$ has infinitely many continuous derivatives in $L^2(\mathbb{R})$, then
	$c_k = O(|k|^{-m})$ as $|k| \to \infty$ for every $m > 0$.
\end{enumerate}
Furthermore, the Paley-Wiener theorem \citep{PaleyWiener:1934, Katznelson:1976} indicates that we should expect an exponential convergence for analytic functions. These results guarantee a very fast convergence for sufficiently smooth solutions.

\subsection{Domain doubling}
One of the features of Chebyshev collocation pseudo-spectral representations is that the point of CGL grid cluster around the boundaries. We consider the distance between two neighboring CGL points \eqref{Num:CGL_grid}
\begin{equation}
	\Delta x_i = x_i - x_{i+1} = 2\sin\left(\frac{\pi}{2N}\right)\left(\frac{\pi(2k + 1)}{2N}\right).
\end{equation}
In the limit of large number of points $N$ the spacing in the center of the domain where $k \sim N/2$ scales as $\Delta x \sim 1/N$, while at the boundary $k\sim 0$ the distance scales as $\Delta x \sim 1/N^2$. This scaling becomes a problem if the equations are singular at the boundary. The situation is especially dire in the spherical coordinates. For example, the Laplacian of a function $f$ in spherical coordinates $(r,\theta,\varphi)$ has the form
\begin{equation}
	\label{Num:Laplacian}
	\Delta f = \frac{1}{r^2}\partial_r\left(r^2 \partial_r f\right) + \frac{1}{r^2\sin\theta}\partial_\theta(\sin\theta\partial_\theta f) + \frac{1}{r^2 \sin^2\theta}\partial_\varphi^2 f.
\end{equation}
If the function $f$ has a non-trivial azimuthal angle $\varphi$ dependence then in the vicinity of $r = 0$ and $\theta = 0$
$$
\frac{1}{r^2 \sin^2\theta} \sim \frac{1}{\varepsilon^4},
$$
where $\varepsilon \sim r,\theta$ is a small quantity of order $r$ and $\theta$. In such situation the code might loose convergence even for moderate domain sizes.

We use the method suggested by \citep{Fornberg:1995, Fornberg:1996, FornbergMerrill:1997, Trefethen:2000}. In many cases sufficiently symmetric solutions in appropriate coordinate systems are odd or even functions of the coordinates. For instance, let us extend the spherical coordinate system into the range of negative values of the radius $r$, the most natural way to do so would be to identify points $(r,\theta,\varphi)$ and $(-r,\pi-\theta,\varphi+\pi)$. If the solution $f$ is an eigenvector of the azimuthal part of the Laplacian and an even function with respect to equatorial plane reflections
$$
f(r,\theta,\varphi) = f(r,\theta) e^{i m \varphi},\quad f(r,\theta) = f(r, -\theta),
$$
this identification would make $f(r,\theta)$ an odd function of radius $f(r,\theta) = - f(-r,\theta)$  for odd $m$ and an even function of radius $f(r,\theta) = f(-r,\theta)$ for even $m$. Similar continuation can be established for the polar angle $\theta$.

In such case we can identify the point $r = 0$ with the center of the domain where the CGL grid is sparse. If we also take the number of CGL points to be even, the point $r = 0$ itself will not be a part of CGL grid \eqref{Num:CGL_grid}. We will call the set of points $x_i$ where $i \in 0,\dots,N-1$ (or alternatively $i \in N,\dots,2N-1$ depending on the doubling side) of the CGL grid
\begin{equation}
	\label{Num:CGL_doubled_grid}
	x_i = \cos\left(\frac{\pi k}{2N - 1}\right)
\end{equation}
the points of the doubled grid.

We can then use the parity of the function $f$ in the collocation expressions (\ref{Num:resum_expansion},\ref{Num:diff_formula}) to  replace the values $f_l$ at the points $r < 0$ with the values at the points $r > 0$, for instance, we can rewrite the derivative formula \eqref{Num:diff_formula} as
\begin{equation}
	f^{(n)}_j = \sum\limits_{l = 0}^{N-1} \left((D^{(n)}_{2N-1})_{j,l} \pm (D^{(n)}_{2N-1})_{j,2N - 1 - l}\right) f_l,
\end{equation}
where the positive sign is taken if the function $f$ is even and the negative sign is taken if the function $f$ is odd, the $N$ here is the number of CGL points in the region $r > 0$. This formulation already contains boundary conditions implicitly so no further boundary conditions must be imposed. In case of multiple functions we require all the functions to have either odd or even parity so they can be set jointly on the same CGL grid.

\section{Differential equation solver}
In this section we summarize the numerical method we use to solve the equations. The method applies to the system of any number of dimensions $d$, although it becomes rather impractical when the number of dimensions is $d > 3$ since the number of collocation points becomes prohibitively high. We will demonstrate how it works in $d = 2$ case of rotating boson stars, however we have implemented the code in such way that it takes the number of dimensions $d$ as a parameter and tested it in $d = 1,2,3,4$

We specify a set of $N_f$ unknown functions $f_r$ that we are going to form into a vector $\vec{f}$ that obey a set of $N_f$ second order elliptic differential equations
\begin{equation}
	\label{Num:unspecified_equations}
	\vec{F}\left[\frac{\partial^2 \vec{f}}{\partial x^a \partial x^b}, \frac{\partial \vec{f}}{\partial x^a}, \vec{f}, x^a\right] = 0.
\end{equation}
We supplement this set of equations with a set of $N_f$ boundary conditions
\begin{equation}
	\label{Num:unspecified_boundary_conditions}
	\vec{G}\left[\frac{\partial \vec{f}}{\partial x^a}, \vec{f}\right] = 0.
\end{equation}
Finally, we also specify parity conditions for all the functions $\vec{f}$ with respect to some of the coordinates $x^a$.

\subsection{Domain decomposition}
We start by defining a computational domain as a $d$-dimensional rectangular block
\begin{equation}
	\mathcal{D} = [a_1, b_1] \times [a_2, b_2] \times \dots \times [a_d, b_d],
\end{equation}
this domain is decomposed into a series of $n_d$ subblocks
\begin{equation}
	\mathcal{D}_p = [a^p_1, b^p_1] \times [a^p_2, b^p_2] \times \dots \times [a^p_d, b^p_d],
\end{equation}
in such way that the domain $\mathcal{D}$ is a union of its subblocks
\begin{equation}
	\mathcal{D} = \bigcup\limits_{p = 1}^{n_d} \mathcal{D}_p,
\end{equation}
and if some point $p$ belongs to two or more subblocks, this point must lie on the boundary of every subblock it belongs to. We will assume that every subblock has nonzero volume or, equivalently, that no $a^p_m$ and $b^p_m$ coincide, where $m$ denotes the $m$-th dimension. We shall also use the term subdomain interchangeably with the term subblock.

The computational domain $\mathcal{D}$ and every subblock $\mathcal{D}_p$ have $(d-1)$-dimensional sides. We introduce the following notation. We call
\begin{equation}
	\mathcal{B}^{(l),p}_m = [a^p_1, b^p_1] \times \dots \times a_m \times \dots \times [a^p_d, b^p_d],
\end{equation}
the left side of the domain $p$, the $m$-th coordinate of any point belonging to this side is fixed to be $a_m$. Similarly we define the right side $\mathcal{B}^{(r),p}_m$ such that the $m$-th coordinate of such side is fixed to be $b_m$. In total, every block has $d$ left sides and $d$ right sides. We will also call the sides the domain boundaries.

On each of the subdomains we define a multidimensional CGL grid, the grid points can be labeled as
\begin{equation}
	x_{i_1\dots i_d} = \left(x^{(1)}_{i_1},\dots,x^{(d)}_{i_d}\right),
\end{equation}
where $x^{(m)}_i$ is a point in $m$-th dimension, we denote the number of points of the subblock $p$ in each dimension $m$ by $n^{(p)}_m$. The total number of points in each subdomain therefore is
\begin{equation}
	N_p = \prod\limits_{i = 1}^d n^{(p)}_i.
\end{equation}

\begin{figure}[H]
	\centering
	\subfloat{\includegraphics[width=0.46\linewidth]{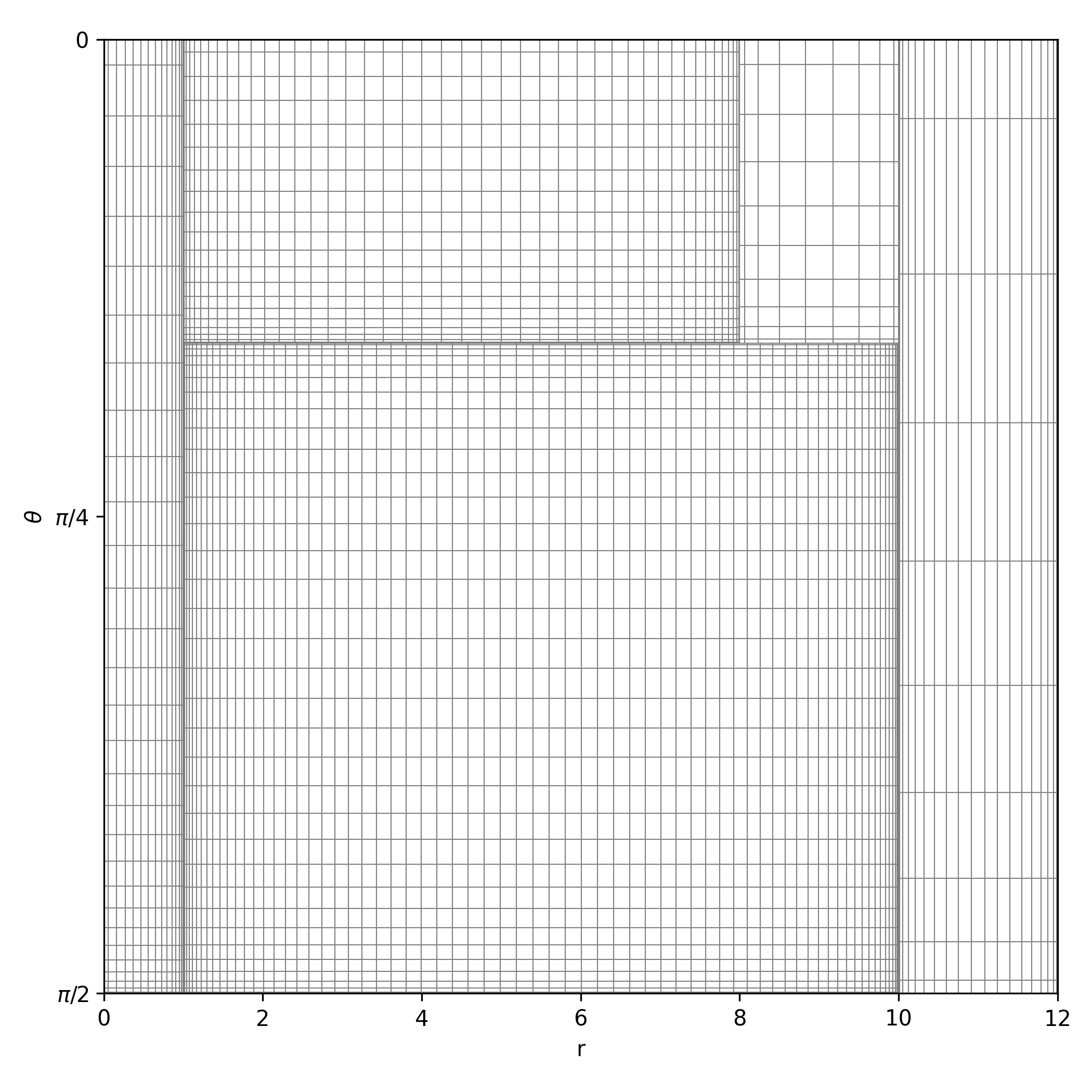}\centering}
	\hfill
	\subfloat{\includegraphics[width=0.50\linewidth]{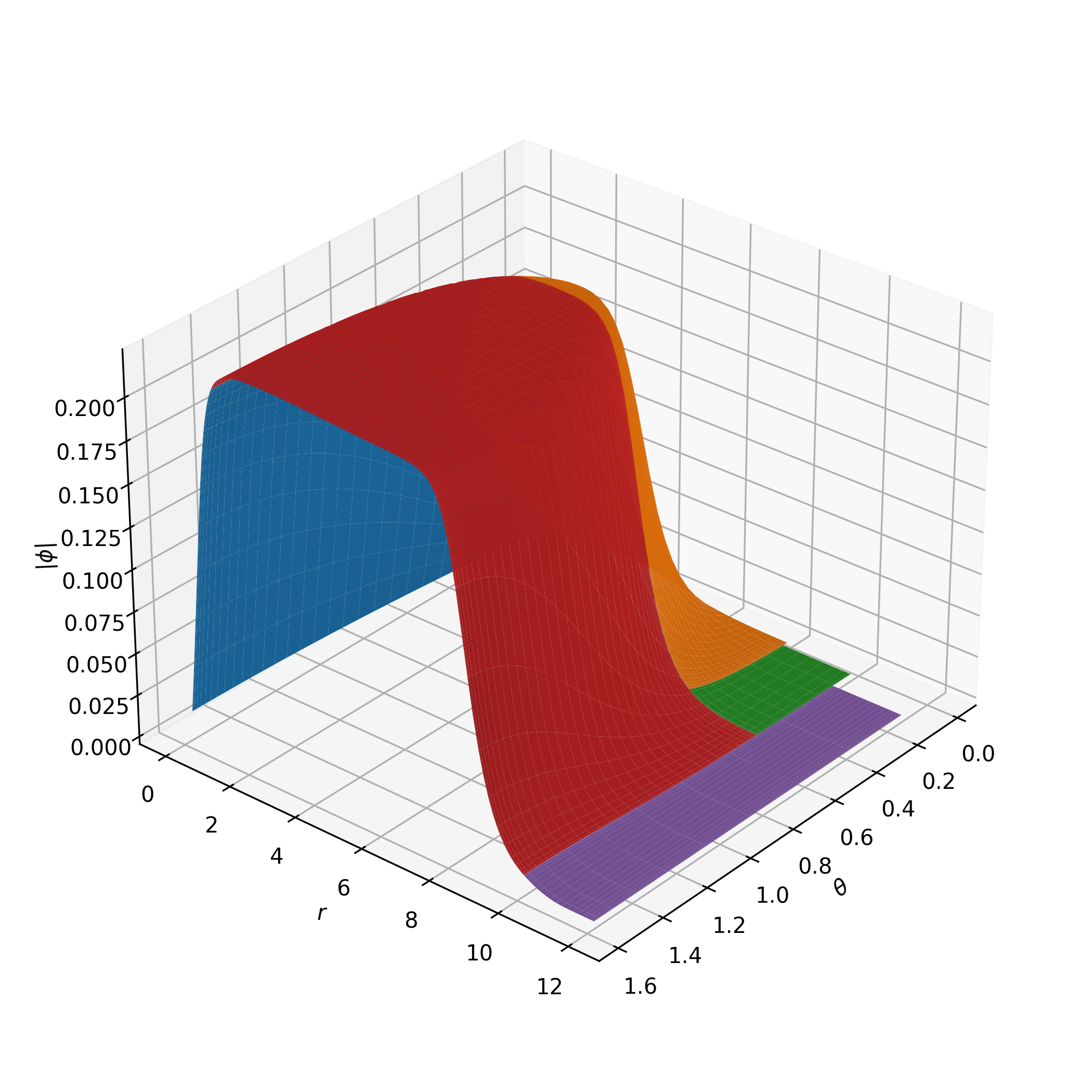}\centering}
	\caption{An example of the domain decomposition together with a function set on the computational domain. Here the function $|\phi|$ is set to be odd both in $r$ and $\theta$ coordinates, the domains adjacent to $r = 0$ and $\theta = 0$ are doubled.}
	\label{Fig:domain_decomposition}
\end{figure}

We can implement the domain doubling along some boundaries $\mathcal{B}_m$ of the computational domain $\mathcal{D}$. In such case the CGL points $x^{(m)}_{i_m}$ along the dimension $m$ of all subdomains $\mathcal{D}_p$ adjacent to the boundary $\mathcal{B}_m$ belong to the doubled grid \eqref{Num:CGL_doubled_grid}. An example of the domain decomposition of a $2D$ computational domain and a function set on such domain is given on the Figure~\ref{Fig:domain_decomposition}.

It might happen that some grid points of two adjacent subdomains coincide, even though this is not a common situation. Tracking down and identifying such points is very cumbersome and doesn't decrease the number of points by much so for all the numerical purposes except the trivial 1D case we will not identify such grid points and guarantee solution matching by an appropriately set system of interdomain matching conditions that we are going to specify later.

\subsection{Setting up equations}
We separate all the grid points of all subdomains of the computational domain $\mathcal{D}$ into three classes
\begin{enumerate}
	\item The points that lie on the boundaries $\mathcal{B}_m$ of the domain $\mathcal{D}$, we call these points the boundary points. We use these points to impose the boundary conditions \eqref{Num:unspecified_boundary_conditions}. The boundaries where the domain doubling was implemented have no grid points so they do not contribute to this class.
	\item The points that lie on the boundaries between subdomains, we call these interdomain points. We will have to specify interdomain matcing conditions at these points.
	\item The rest of the points that lie in the bulks of the subdomains, we call these the bulk points. We use these points to impose the equations \eqref{Num:unspecified_equations}. None of these points lie on the boundaries so none of the equations \eqref{Num:unspecified_equations} are singular at these points even if some equations are singular at the boundaries.
\end{enumerate}
To set up an equation or a boundary condition at any gridpoint $x_{i_1\dots i_d}$ we use the differentiation formula \eqref{Num:diff_formula} independently along each axis. This way we convert the set of differential equations and boundary conditions in each subdomain into a set of algebraic equations for the values of the functions $\vec{f}$ at the collocation points of the subdomain.

Now suppose that we have a set of functions $\vec{f}^{(p)}$ that are solutions to the equations \eqref{Num:unspecified_equations} and the boundary conditions \eqref{Num:unspecified_boundary_conditions} on the subdomains $\mathcal{D}_p$. We need to ensure that these functions represent the same solution on the computational domain $\mathcal{D}$ as a whole. The equations \eqref{Num:unspecified_equations} are of the second order, if we have two subdomains $\mathcal{D}_p$ and $\mathcal{D}_q$ that share a common boundary
\begin{equation}
	\hat{\mathcal{B}} = \mathcal{B}^{(r), p}_m \cap \mathcal{B}^{(l), q}_m \neq \varnothing,
\end{equation}
where $m$ is the dimension of the normal vector to the boundary as we defined earlier, we need to ensure that the values and the normal derivatives of $\vec{f}^{(p)}$ and $\vec{f}^{(q)}$ match on the common boundary $\hat{\mathcal{B}}$, or
\begin{equation}
	\label{Num:interdomain_conditions}
	\left.\vec{f}^{(p)}\right|_{\hat{\mathcal{B}}} = \left.\vec{f}^{(q)}\right|_{\hat{\mathcal{B}}},\quad \left.\partial_{x^m} \vec{f}^{(p)}\right|_{\hat{\mathcal{B}}} = \left.\partial_{x^m} \vec{f}^{(q)}\right|_{\hat{\mathcal{B}}}.
\end{equation}
As we mentioned earlier, in general we do not expect the collocation points to match so to impose these conditions we use the formulas \eqref{Num:resum_expansion} that allow us to interpolate the functions set on CGL grid to an arbitraty domain point. There is a number of different equivalent ways of setting the matching conditions, we are going to do the following two stage process:
\begin{enumerate}
	\item We take all the gridpoints on the right boundary $\mathcal{B}^{(r), p}_m$ of the subdomain $\mathcal{D}_p$ and use the expression \eqref{Num:resum_expansion} to interpolate the functions $\vec{f}^{(p)}$ to the gridpoints of the subdomain $\mathcal{D}_q$ that lie on the intersection $\hat{\mathcal{B}}$. We then demand that these functions match the functions $\vec{f}^{(q)}$ at these points.
	\item We use the differentiation matrices \eqref{Num:diff_formula} to find the derivatives $\partial_{x^m} \vec{f}^{(q)}$ at the left boundary $\mathcal{B}^{(l), q}_m$, we then use the interpolation expressions \eqref{Num:resum_expansion} to interpolate the derivatives to the $\mathcal{D}_p$ gridpoints that lie on the intersection $\hat{\mathcal{B}}$. We again use the differentiation matrices \eqref{Num:diff_formula} to find $\partial_{x^m} \vec{f}^{(p)}$ derivatives at the subdomain $\mathcal{D}_p$ gridpoints that lie on the intersection $\hat{\mathcal{B}}$ and demand that these values are equal to the interpolated $\partial_{x^m} \vec{f}^{(q)}$ values.
\end{enumerate}
The resulting matching conditions are a set of linear equations in terms of the collocation values $\vec{f}^{(p)}_l$ and $\vec{f}^{(q)}_l$ of the functions $\vec{f}^{(p)}$ and $\vec{f}^{(q)}$ set on the collocation grids of the subdomains $\mathcal{D}_p$ and $\mathcal{D}_q$. The number of these equations in total matches the total number of the gridpoints that lie on interdomain boundaries by construction. There is one last ambiguity we have to resolve, in all dimensions $d \geq 2$ there are points that lie on multiple $(d-1)$-dimensional sides of the $d$-dimensional subdomains (e.g. in $d = 2$ case as in Figure~\ref{Fig:domain_decomposition} these are the points that lie in the corners of the rectangles), at such points we have to select the direction of the normal derivative we should take. It doesn't matter much in principle what we choose so we are going to choose the dimension with the highest index in $x_{i_1\dots i_d}$ that is orthogonal to the side intersection.

\subsubsection{Compactification}
Some of the coordinates may extend up to infinity, is such cases we implement compactification. For example, for the radial coordinate $r$ in spherical coordinates we adopt the compactified coordinate
\begin{equation}
	\label{Num:compactification}
	\bar{r} = \frac{r}{1 + r},
\end{equation}
the interval $\bar{r} \in [0,1)$ covers the whole radial coordinate range $r\in [0,\infty)$. We found that it is best to only use compactification in the subdomains that touch the spatial infinity. There are two reasons for that. The first reason is that we need a special choice of compactification to preserve parity, for instance if a function $f(r)$ is an odd or an even function, the function $f(\bar{r})$ of the compactified coordinate $\bar{r}$ according to \eqref{Num:compactification} will be neither odd nor even. The second reason is that we empirically find that the code converges much faster when only few domains use compactification.

\subsection{Solver iteration step}
The method to set up the equations that was described in the previous section specifies $N_f$ algebraic equations at each collocation point. These equations are independent and the number of the unknown functions $\vec{f}$ matches the number of the equations \eqref{Num:unspecified_equations}, thus the number of the equations in total matches the total number of the unknown values $\vec{f}_l$ at the gridpoints. We find that we don't need any preconditioning as opposed to what is suggested by \citep{Pfeiffer:2002wt}.

Let us call $\mathbf{f}$ a large vector $\vec{f}_l$ that consists of all $N_f$ functions at all grid points of all subdomains of the computational domain and $\mathbf{S}$ the vector made of all the equations at all gridpoints of the computational domain. The discretized version of the equations \eqref{Num:unspecified_equations} together with the boundary conditions \eqref{Num:unspecified_boundary_conditions} and interdomain matching conditions can then be expressed as
\begin{equation}
	\label{Num:total_system}
	\mathbf{S}[\mathbf{f}] = 0.
\end{equation}
The interdomain matching equations are always linear in $\vec{f}_l$, however, if the equations \eqref{Num:unspecified_equations} are nonlinear the resulting algebraic equations set at the bulk points are nonlinear as well. Therefore in general the operator $\mathbf{S}$ is nonlinear.

We use the Newton-Raphson method with line searches as described in \citep{PressTeukolsky:2007} to solve the system \eqref{Num:total_system}. The Newton-Raphson step requires the knowledge of both the equation $\mathbf{S}[\mathbf{f}]$ guess solution as well as the Jacobian matrix
\begin{equation}
	\label{Num:Jacobian}
	\mathbf{J} \equiv \frac{\partial S}{\partial \mathbf{f}}[\mathbf{f}_{old}].
\end{equation}

The computation of $\mathbf{S}[\mathbf{f}]$ is fast and can be easily parallelized. The Jacobian \eqref{Num:Jacobian} can be computed via the chain rule
\begin{equation}
	\mathbf{J}_{ik} = \sum_l\left[\frac{\partial S_i}{\partial \vec{f}_k} + \frac{\partial S_i}{\partial (\partial\vec{f}_l)} D^{(1)}_{lk} + \frac{\partial S_i}{\partial (\partial^2 \vec{f}_l)} D^{(2)}_{lk}\right],
\end{equation}
where indices $i,k,l$ correspond to different collocation points and $D^{(1)}$ and $D^{(2)}$ are differentiation matrices defined by expressions \eqref{Num:diff_matrix_first} and \eqref{Num:diff_matrix_second}. The chain rule is symbolic, we don't carefully track the different subdomain indices and dimension indices since it would make the expression very cumbersome and is irrelevant to the point. We could compute the derivatives of the equations \eqref{Num:unspecified_equations} with respect to the functions $\vec{f}$ and the derivatives  $\partial \vec{f}/\partial x^a$, $\partial^2 \vec{f}/(\partial x^a \partial x^b)$ at each grid point analytically, however, for convenience we instead take numerical derivatives
\begin{equation}
	\frac{\partial F_s}{\partial f_r} \approx \frac{F_s(f_r + h) - F_s(f_r - h)}{2h},
\end{equation}
where $h$ is a small constant. After computing equations' derivative at a gridpoint it becomes very easy to compute the whole rows of the Jacobian via the chain rule since the differentiation matrices are known explicitly. Similarly to $\mathbf{S}[\mathbf{f}]$ the computation of Jacobian is fast and can be easily parallelized.

The main computational bottleneck of the Newton-Raphson method lies in solving the system of linear equations for the step update vector. Indeed a single $2D$ computational domain with $N_x \times N_y$ gridpoints and $N_f$ functions will have the solution vector $\mathbf{f}$ of the size $N_x N_y N_f$. The size of the corresponding Jacobian is $N_x^2 N_y^2 N_f^2$. In practice, however, we don't find this scaling to be particularly taxing because linear equations solvers are very well optimized and parallelized, we can selectively put more gridpoints to the subdomains where we need more resolution and put fewer points where the solution is smooth and featureless and because spectral methods in general scale exponentially and we don't need very large numbers of collocation points to achieve good accuracy.

\subsection{Convergence and error control}
To test the convergence of the method we described we consider a rotating BS system with the solitonic potential. The system is described by the action
\begin{equation}
	\label{Description:Action}
	S = \int d^4x \sqrt{-g} \left[\frac{R}{2\kappa}- g^{ab} \partial_a \Phi^* \partial_b \Phi - V(|\Phi|^2)\right],
\end{equation}
where $\Phi$ is the complex scalar field with the potential $V(|\Phi|^2)$ given by the expression \eqref{Description:Solitonic_potential}, $g$ is the metric determinant and $R$ is the Ricci scalar. Varying the action with respect to the field and the metric we obtain the system of field equations
\begin{subequations}
	\label{Description:star_equations}
	\begin{align}
		\label{Description:Einstein_equations}
		R_{ab} - \frac{1}{2} R g_{ab} = \kappa T_{ab},\\
		\label{Description:KG_equation}
		\square\Phi - V'(|\Phi|^2)\Phi = 0,
	\end{align}
\end{subequations}
where $T_{ab}$ is the canonical stress-energy tensor
\begin{equation}
	\label{Description:stress-energy tensor}
	T_{ab} = \partial_a \Phi \partial_b \Phi^* + \partial_a \Phi^* \partial_b \Phi - g_{ab}\left[g^{cd} \partial_c \Phi \partial_d \Phi^* + V(|\Phi|^2)\right].
\end{equation}
To describe the spacetime geometry we pick the metric ansatz
\begin{equation}
	\label{Description:KEH_ansatz}
	ds^2 = - e^{\gamma + \rho}dt^2 + e^{2\sigma}(dr^2 + r^2 d\theta^2) + e^{\gamma - \rho} r^2 \sin^2\theta (d\varphi - \omega dt)^2,
\end{equation}
given in spherical-like coordinates $(t,r,\theta,\varphi)$ where the metric functions $\gamma(r,\theta)$, $\rho(r,\theta)$, $\omega(r,\theta)$ and $\sigma(r,\theta)$ are the functions of the radial coordinate $r$ and the polar angle $\theta$ alone. The scalar field profile for a rotating BS in the coordinate system \eqref{Description:KEH_ansatz} is a simple harmonic function in time $t$ and azimuthal angle $\varphi$
\begin{equation}
	\label{Description:phi_ansatz}
	\Phi = \phi(r,\theta)\,  e^{i \Omega t + i m \varphi},
\end{equation}
where $\Omega \in \mathbb{R}$ is the scalar field oscillation frequency and $m \in \mathbb{N}$ is \textit{the azimuthal harmonic index} or \textit{the azimuthal winding number}, this number is intimately related to the star angular momentum. We will focus on the main branch $m = 1$. Due to the stress-energy conservation $\triangledown_a T^{ab} = 0$ there are two excess equations that are zero provided that the rest of the equations are satisfied and can be used as independent residuals. The explicit system of equations as well as the exact choice of independent residuals are given in the Appendix. We are free to choose the length scale and the energy scale of the system which allows us to set $\kappa = 1$ and $\mu = 1$. This choice effectively leaves $\sigma_0$ as the sole independent action parameter. \footnote{The other popular choice $\kappa = 8\pi$ will lead to a different normalization for $\sigma_0$, in particular $\left.\sigma_0\right|_{\kappa = 1} = \sqrt{8\pi} \left.\sigma_0\right|_{\kappa = 8\pi}$.}

\begin{figure}[H]
	\centering
	\subfloat{\includegraphics[width=0.46\linewidth]{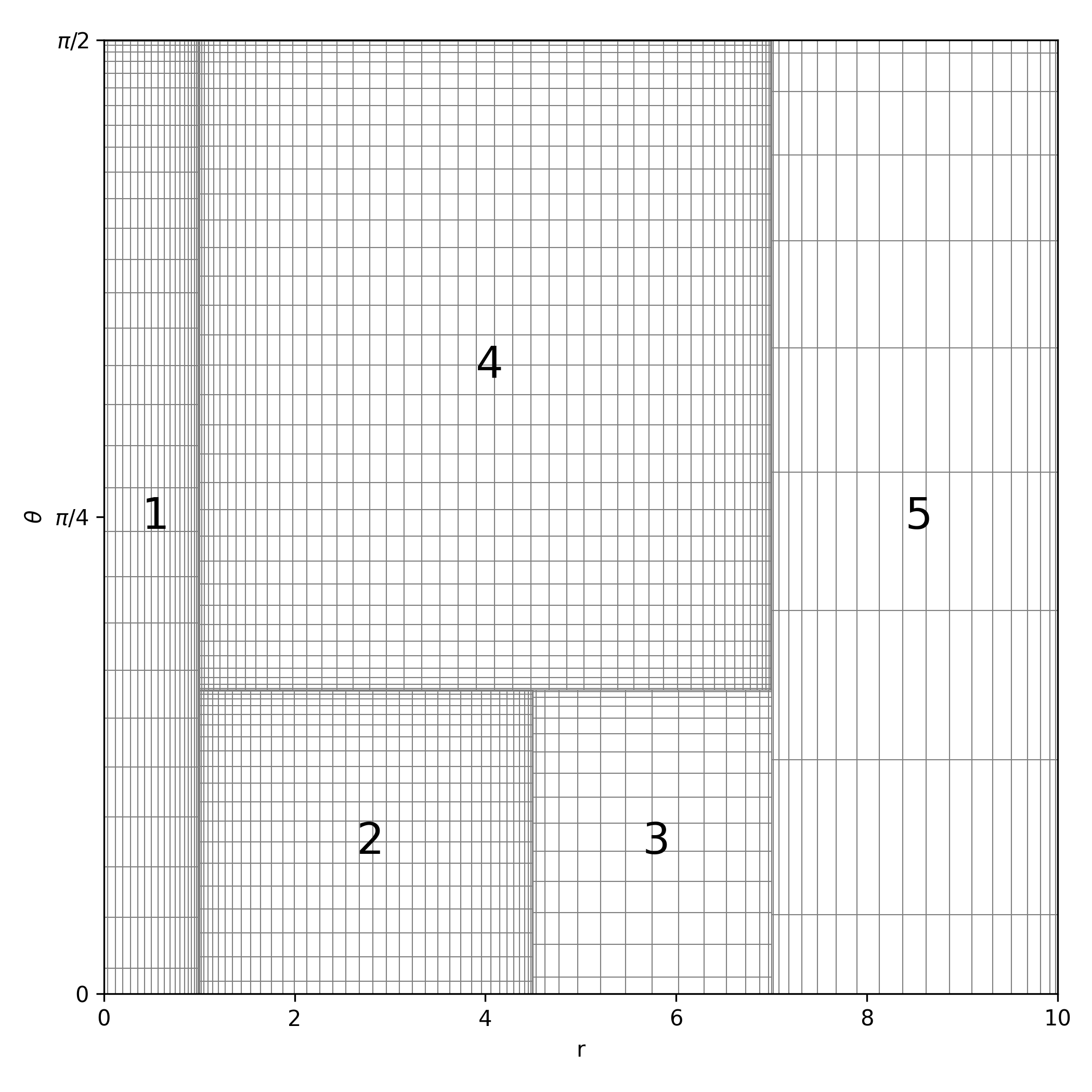}\centering}
	\hfill
	\subfloat{\includegraphics[width=0.5\linewidth]{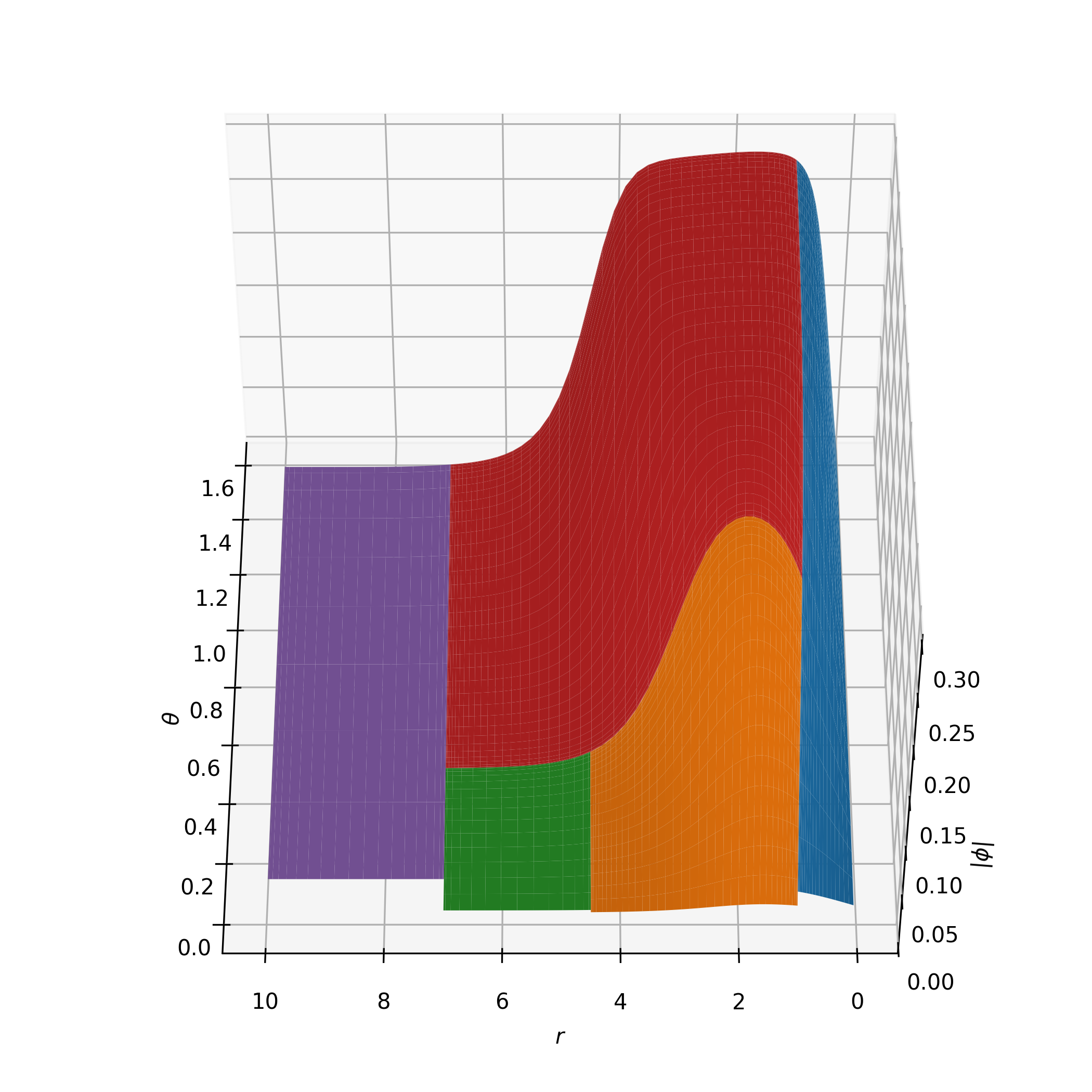}\centering}
	\caption{A solitonic solution (with the potential parameter $\sigma_0 = 0.4$) set on the computational domain that was split into five subdomains. The left figure represents the domain decomposition, the right figure demostrates the absolute value of the scalar field $\Phi$ set on the domain decomposition. The radial coordinate in the subdomain 5 is compactified with the transformation \eqref{Num:compactification}, $r = 10$ corresponds to the spatial infinity.}
	\label{Fig:solution_convergence}
\end{figure}

\begin{figure}[H]
	\begin{center}
		\includegraphics[width=1\linewidth]{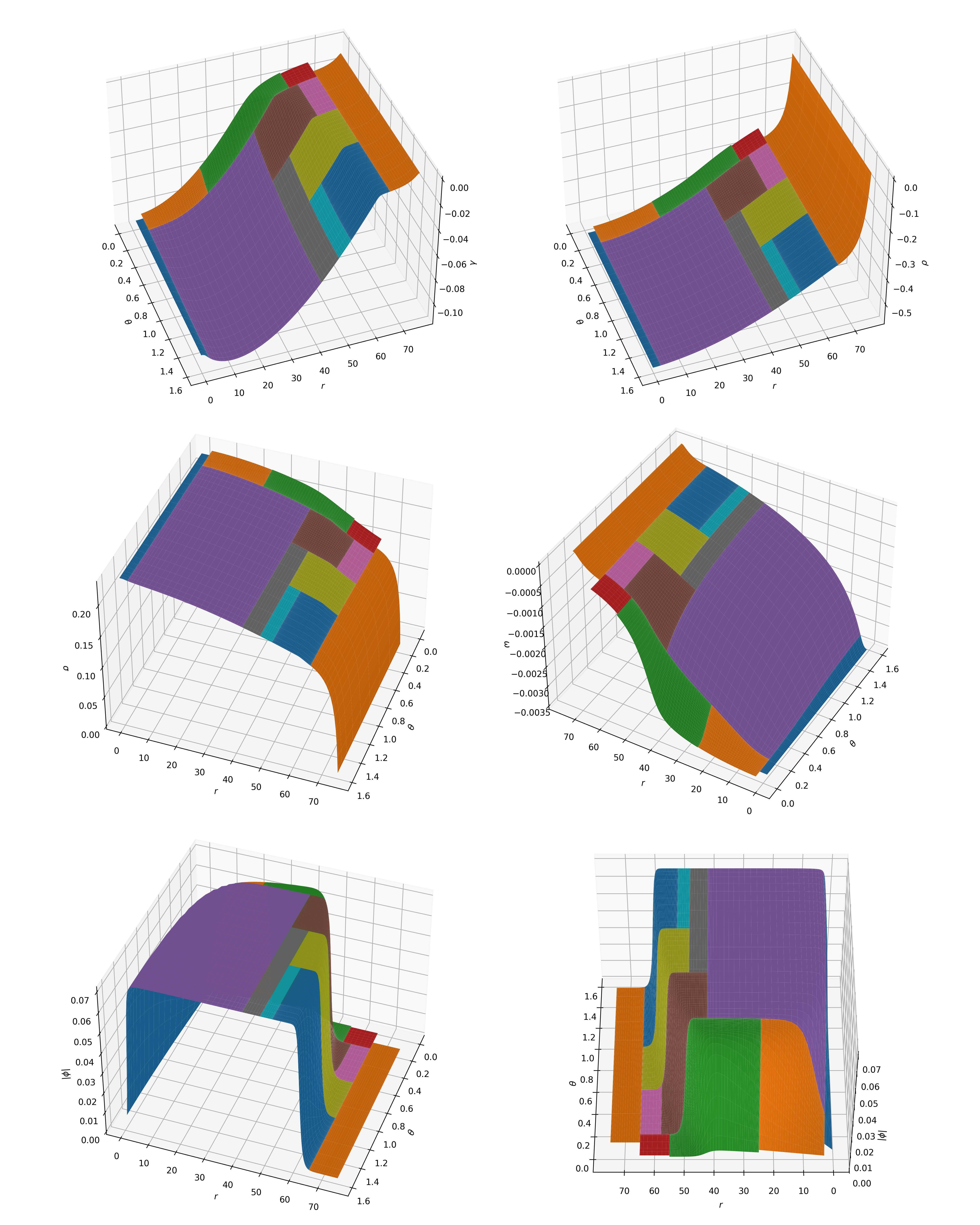}
	\end{center}
	\caption{The metric functions and the scalar field profile of a very stiff solitonic star solution (with the potential parameter $\sigma_0 = 0.1$). As before, the radial compactification is only present in the domain touching the spatial infinity.}
	\label{Fig:stiff_solution}
\end{figure}

The independent residuals can be used as an estimate of the numerical error, the numerical deviation of the independent residuals from zero serves as a good indicator of the subdomains and locations which require resolution increase. To illustrate the idea we take a rotating solitonic boson star solution on Figure~\ref{Fig:solution_convergence}. We typically find that the residuals are the highest in the regions where the rate of change of the solution as compared to the grid spacing is the largest. For solitonic boson stars the metric functions are typically much smoother than the scalar field and the sharpest changes occur in the scalar field (the extreme example of this situation is depicted on Figure~\ref{Fig:stiff_solution}).

\begin{figure}[H]
	\includegraphics[width=1\linewidth]{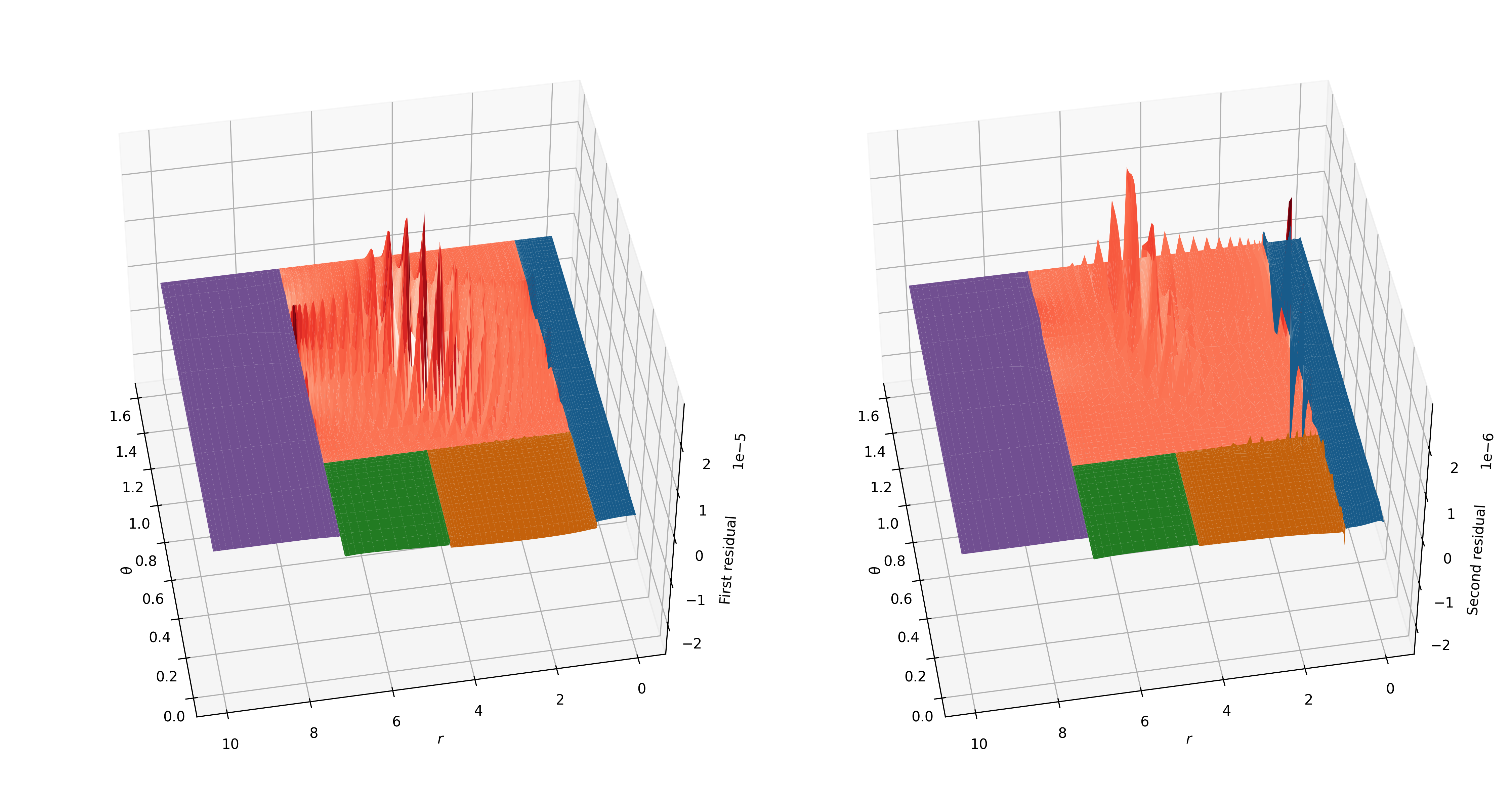}
	\caption{The values of the first \eqref{Model:first_residual} and the second \eqref{Model:second_residual} independent residuals. We see that the region with the highest values of residuals corresponds to the region of the fastest scalar field change on the Figure~\ref{Fig:solution_convergence}.}
	\label{Fig:residual_plots}
\end{figure}

The first \eqref{Model:first_residual} and the second \eqref{Model:second_residual} residuals are shown on the Figure~\ref{Fig:residual_plots}, we see that the locations of residuals maxima correspond to the region where the scalar field experiences the most rapid change.

We use the $L_\infty$ norm to estimate the overall numerical error, which means we take the highest overall absolute value of the residuals in the computational domain. Typically both residuals reach their maxima close to each other so for the purpose of this discussion we don't need to consider the residuals separately. The typical pattern of convergence of the multidomain collocation solver we have described is as we increase the resolution of the subdomain with the highest magnitude of independent residuals the maximum of the residuals converges exponentially with the number of gridpoints until the residuals in some other domain start to dominate. Naturally, there is no practical point in increasing the resolution of the former subdomain beyond that. We have also tested the convergence by other means, e.g. the convergence of Komar integral mass and momentum to the mass and momentum derived from the asymptotic, however since the $L_\infty$ residual convergence is the most stringent convergence test we are not including the other test results.

\begin{figure}[H]
	\begin{center}
		\includegraphics[width=0.7\linewidth]{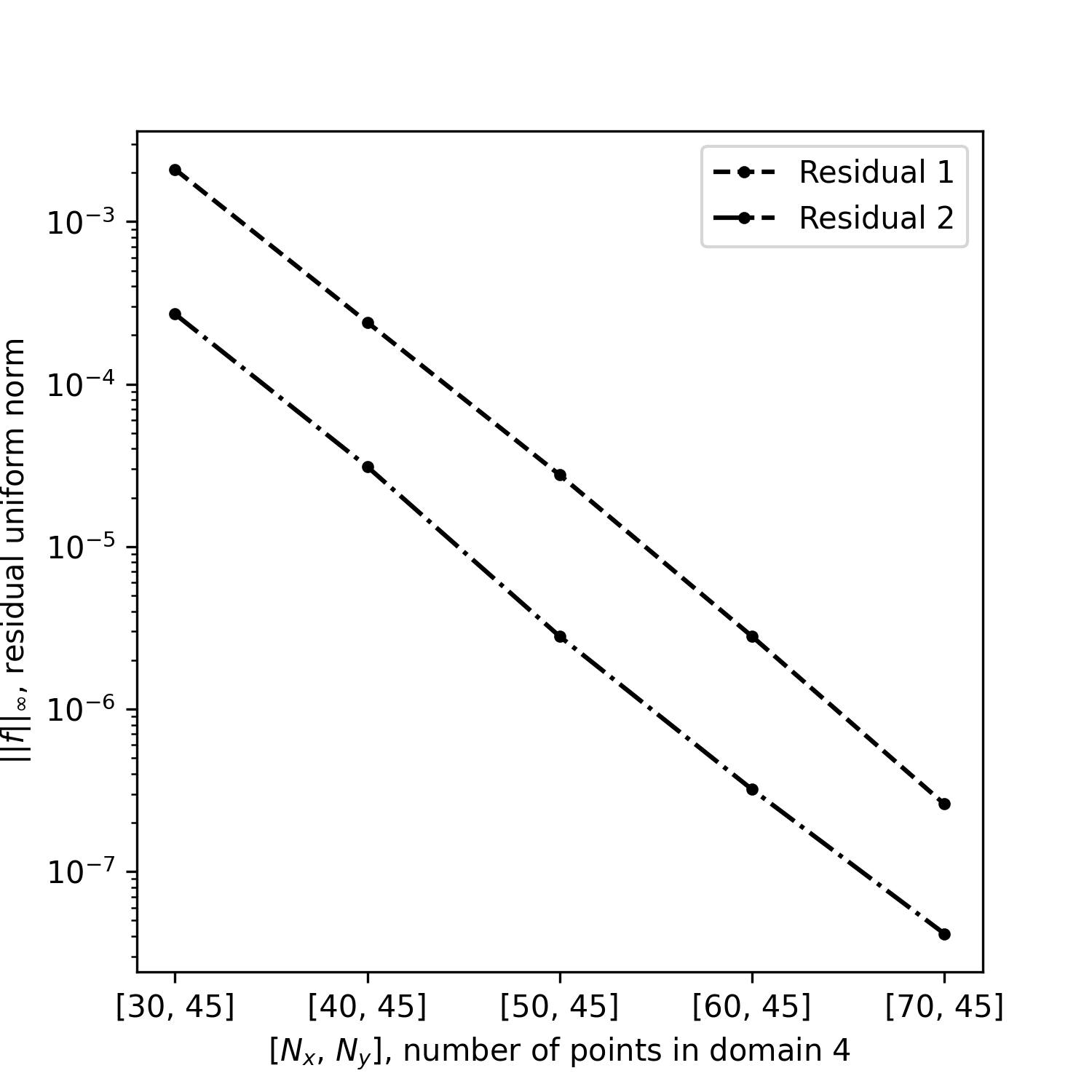}
	\end{center}
	\caption{Convergence of the first \eqref{Model:first_residual} and the second \eqref{Model:second_residual} residuals as we jointly increase the number of collocation points along the coordinate axes in the domain with the highest residuals values.}
	\label{Fig:convergence_plot}
\end{figure}

We demonstrate this behavior on the same solution displayed on the Figure~\ref{Fig:solution_convergence}, we have chosen the number of gridpoints in the subdomains $1,2,3,5$ such that the residuals in these domains are negligible. Figure~\ref{Fig:convergence_plot} demonstrates the exponential convergence as we increase the number of collocation points along both axes.
We increase the number of gridpoints along both axes jointly for demonstration purposes. In practice, the residual errors are usually dominated by the number of points along only one of the axes.

In principle one could use the locations of the maxima of independent residuals in a heuristic adaptive mesh refinement algorithm. However, in practice we found out that for the boson stars it is usually fairly straightforward to do the domain decomposition by hand. For instance we take a very stiff solution demonstrated on Figure~\ref{Fig:stiff_solution}.

\begin{figure}[H]
	\begin{center}
		\includegraphics[width=0.7\linewidth]{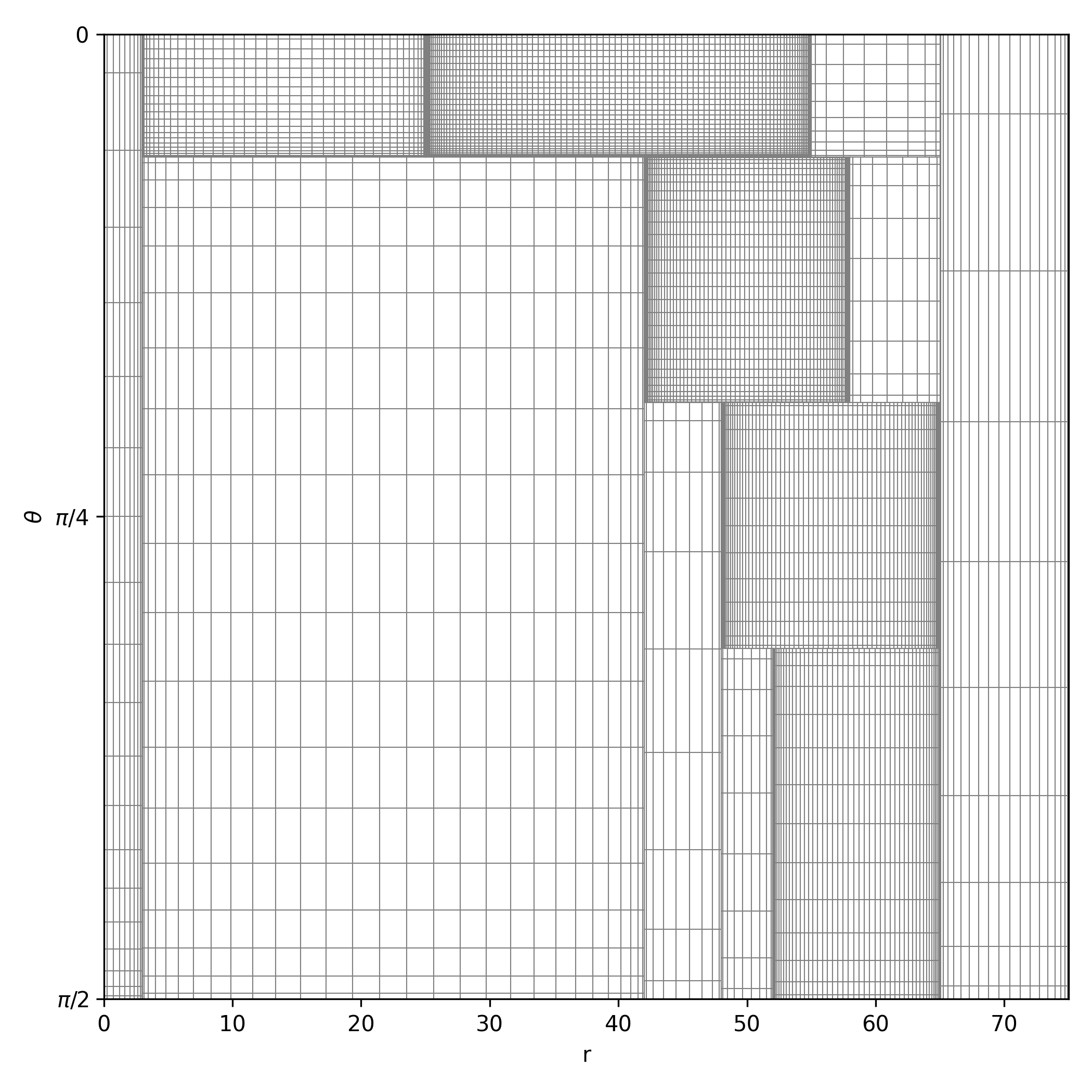}
	\end{center}
	\caption{The scalar field profile for a very stiff solitonic star solution (with the potential parameter $\sigma_0 = 0.1$)}
	\label{Fig:stiff_domain_decomposition}
\end{figure}

To optimally distribute resources we cover the cusp of the solution with subdomains with a very high density of the collocation points while keeping the relatively uniform regions covered with subdomains with few collocation points, we show the domain decomposition of the stiff solution from the Figure~\ref{Fig:stiff_solution} in more detain on Figure~\ref{Fig:stiff_domain_decomposition}.

\section{Non-rotating solitonic boson stars}
As an independent test for our method we study spherically symmetric solitonic boson star solutions. These BS are interesting for a number of reasons; they have a well-known analytical approximation \citep{Friedberg:1986tq} that was tested in the solitonic limit numerically by \citep{Collodel:2022jly}. This happens because, unlike the rotating boson stars with a complicated cusp profile (e.g. the star shown on the Figure~\ref{Fig:stiff_solution}), the stiff part of the solution happens on an interval that can be fitted as a whole into a single domain, and the scalar field solution elsewhere can be taken to be approximately constant. The SBS are also very hard to find by the conventional shooting method \citep{Macedo:2013jja} typically used in spherical symmetry.

\subsection{Solitonic boson star equations in spherical symmetry}
Static boson star equations obey the same system of field equations \eqref{Description:Einstein_equations}, however, the metric and the scalar field configurations are spherically symmetric. To describe the spacetime geometry we adopt Schwarzschild-like coordinates with the metric given by
\begin{equation}
	\label{Description:Schwarzschild metric}
	ds^2 = - e^{v(r)} dt^2 + e^{u(r)} dr^2 + r^2\left(d\theta^2 + \sin^2\theta d\phi^2\right).
\end{equation}
The spherically symmetric scalar field configuration can be described by the expression
\begin{equation}
	\label{1Dstar:Scalar_field_form}
	\phi(t, r) = \phi_0(r) e^{i\Omega t}.
\end{equation}
where $\Omega \in \mathbb{R}$ is the scalar field oscillation frequency. Due to the stress-energy conservation $\triangledown_a T^{ab} = 0$ there is one excess equation that is zero provided that the rest of the equations are satisfied and can be used as an independent residual. The explicit system of equations as well as the independent residual are given in the Appendix.

\subsubsection{Star parameters}
Spherically symmetric spacetimes have much simpler equations that can be integrated analytically. The integral of the equation \eqref{1Dstar:gravity_u_equation} can be expressed as
\begin{equation}
	e^{-u} = 1 - \frac{2M(r)}{r},
\end{equation}
where
\begin{equation}
	\label{1Dstar:MS_mass}
	M(r) = \frac{\kappa}{2} \int\limits_0^r \rho r^2 dr.
\end{equation}
The expression \eqref{1Dstar:MS_mass} is a convenient definition of the quasilocal mass called Misner-Sharp mass. The Misner-Sharp mass \eqref{1Dstar:MS_mass} can be used to define the radius of the boson star. Unlike fluid stars that have a sharp boundary, boson star energy densities decay exponentially at infinity, so we define the radius of the boson star as the radius of the sphere that contains 99\% of the star Misner-Sharp mass. The star also has a conserved U(1) charge given by the expression
\begin{equation}
	Q = i\int n^\mu (\phi\,\partial_\mu \phi^* - \phi^* \partial_\mu \phi) dV = 8 \pi \Omega \int\limits_0^\infty e^{\frac{u - v}{2}} \phi^2 r^2 d r,
\end{equation}
where $n^\mu$ is a unit normal vector to $t = const$ family of hypersurfaces and $dV$ is the hypersurface volume element.

\subsection{Exploring solitonic boson star parameter space}
The spherically symmetric stars corresponding to a specific potential form a series of single parameter families characterized by the number of nodes or points where the scalar field cross zero. We focus on the fundamental family. Typically each family of boson stars have stable and unstable parts. The stability of the stars to radial perturbations follows the logic also found in the fluid star cases \citep{Weinberg:1972kfs}, the transition between stability and instability happens at the extrema of the star mass and charge.However, not every extremum separates a stable intervals from unstable, it can also happen when a few unstable modes are already present and another one is about to become unstable. In this section we will only focus on the parts of the stable sections of solitonic boson star families that cover the most massive and shell-like stars.

The pseudo-spectral Chebyshev multidomain method we described requires an initial guess or a seed as a starting point. It is easy to find a good seed for sufficiently smooth and spread out stars, one could typically use a gaussian shaped scalar field and a flat metric as a starting point. However, the convergence of the Newton-Rahpson algorithm in practice is not guaranteed for every guess, as we progress to stiffer solutions, guessing a good seed becomes hard and impractical.

The method we found works the best is to start with a few solutions in the region of the family parameter curve with sufficiently smooth boson stars and then use these solutions for pointwise extrapolation to a new point on the curve and use the extrapolation result as a seed. This method can then be repeated to continuously go through the whole curve. The stepsize for traversing the curve can be adjusted by looking at the number of iteration the Newton-Raphson method takes to converge: if there were too many iterations on the previous step, the stepsize can be decreased.

To implement the extrapolation idea we need to choose an appropriate parametrization of the curve. Unfortunately, neither of the usual parameters like the star oscillation frequency $\Omega$, the mass $M$, the charge $Q$, the star radius $R$ or the absolute value of the scalar field at the origin $\phi_c$ are single-valued along the whole curve. One could of course choose different extrapolation parameters at different sections of the family curve, but there is a better substitute for this switching. We define the rectified length of a curve as
\begin{equation}
	\label{1DStar:pseudolength}
	L = L_0 + \sum\limits_i \sqrt{\sum\limits_a \left(\frac{p^{i+1}_a - p^i_a}{p^i_a}\right)^2},
\end{equation}
where $a$ denotes the parameter type (e.g. mass or radius) and $i$ is the point number along the curve in the order we obtained it. We have introduced $p_a^{i}$ in denominator to account for different scales of the parameters. We record the parameter \eqref{1DStar:pseudolength} for a first few star solutions and then extrapolate the frequency $\Omega$ and the star profile to some close increased value of the rectified length $L$, solve boson star equations for the obtained guess and then record the rectified length \eqref{1DStar:pseudolength} for the new star. We then repeat this step to obtain the full family parameter curve. In practice, we found this method to be very efficient for extrapolating along the family curve in the star parameter space.

The final comment is that in $1D$ case the boundary of the domain consists of a single point and hence we don't need the complicated domain matching we use in the multidimensional case. Furthermore, we can identify boundary points of the adjacent domains and only impose the derivative matching \eqref{Num:interdomain_conditions} at interdomain boundary points.

\subsection{Spherically symmetric results}
\begin{figure}[H]
	\begin{center}
		\includegraphics[width=1\linewidth]{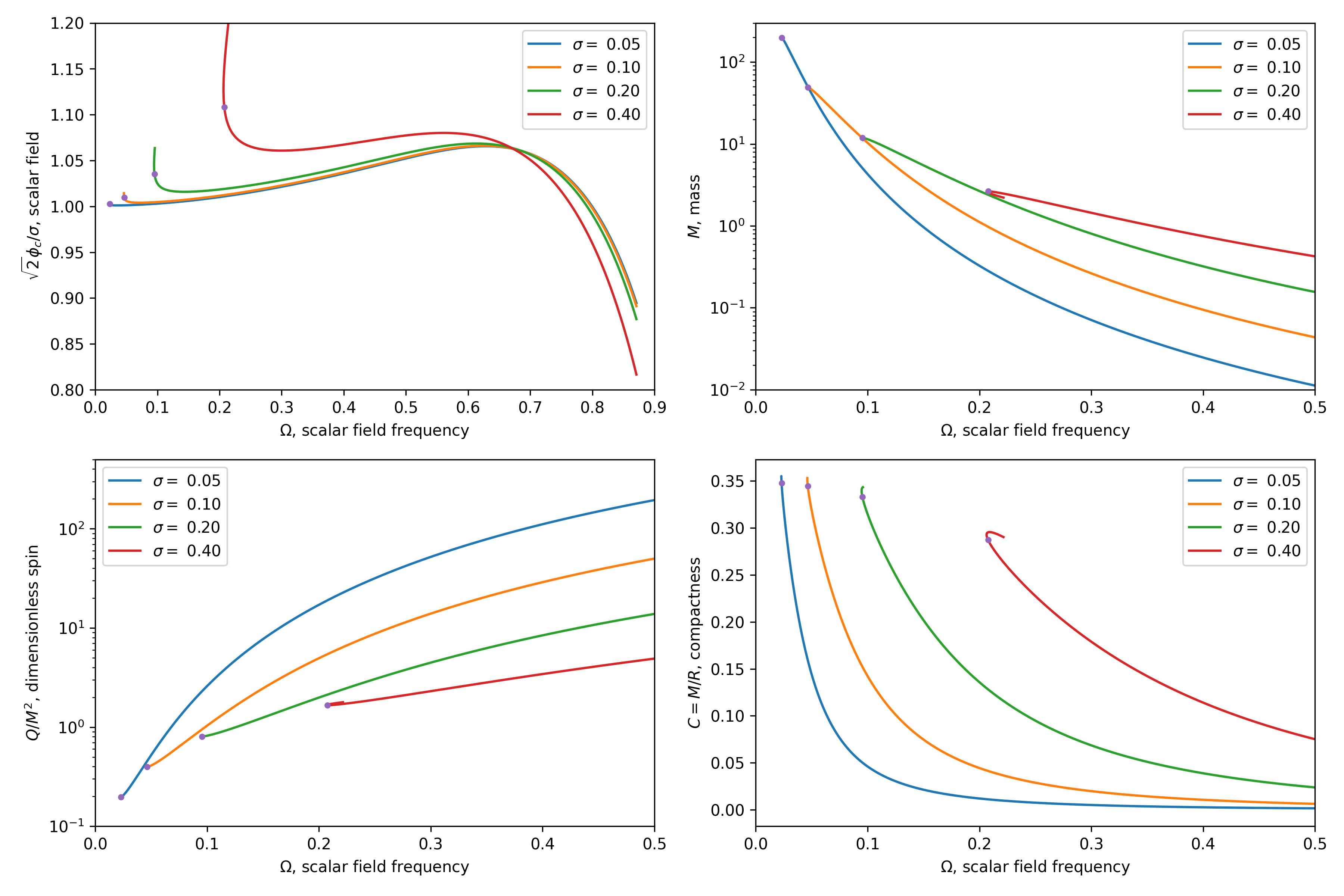}
	\end{center}
	\caption{Solution sets for different values of $\sigma_0$. We have not explored the curves past the mass and the charge maxima since those regions of the family are unstable and thus not very interesting. The purple dots correspond to the mass extrema and thus the ends of the stable parts of BS families.}
	\label{Fig:spherical_star_tracing}
\end{figure}

\begin{figure}[H]
	\begin{center}
		\includegraphics[width=1\linewidth]{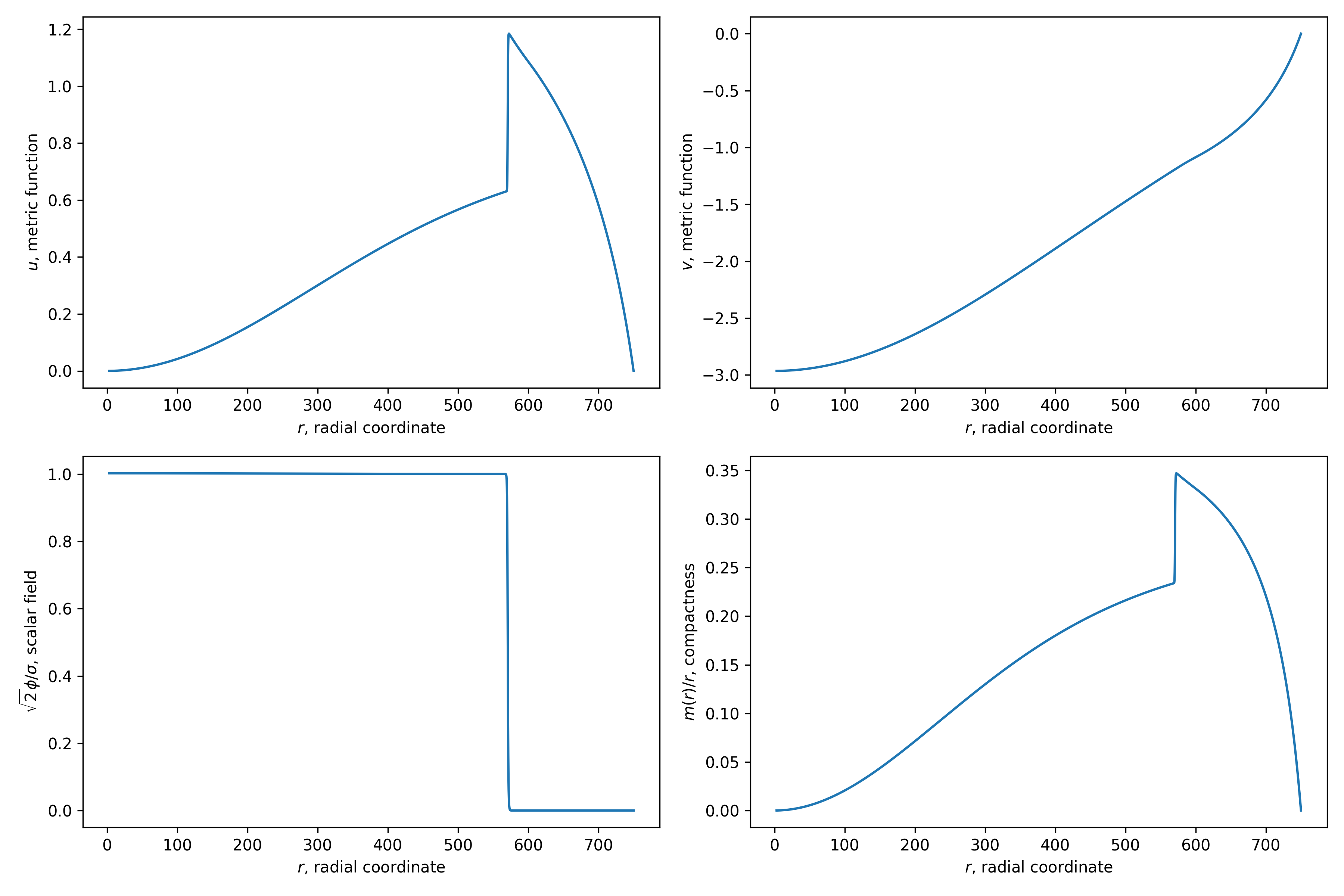}
	\end{center}
	\caption{The scalar field profile for a very stiff solitonic star solution (with the potential parameter $\sigma_0 = 0.05$). The radial compactification \eqref{Num:compactification} starts at $r = 600$.}
	\label{Fig:spherical_star_example}
\end{figure}

\begin{figure}[H]
	\begin{center}
		\includegraphics[width=1\linewidth]{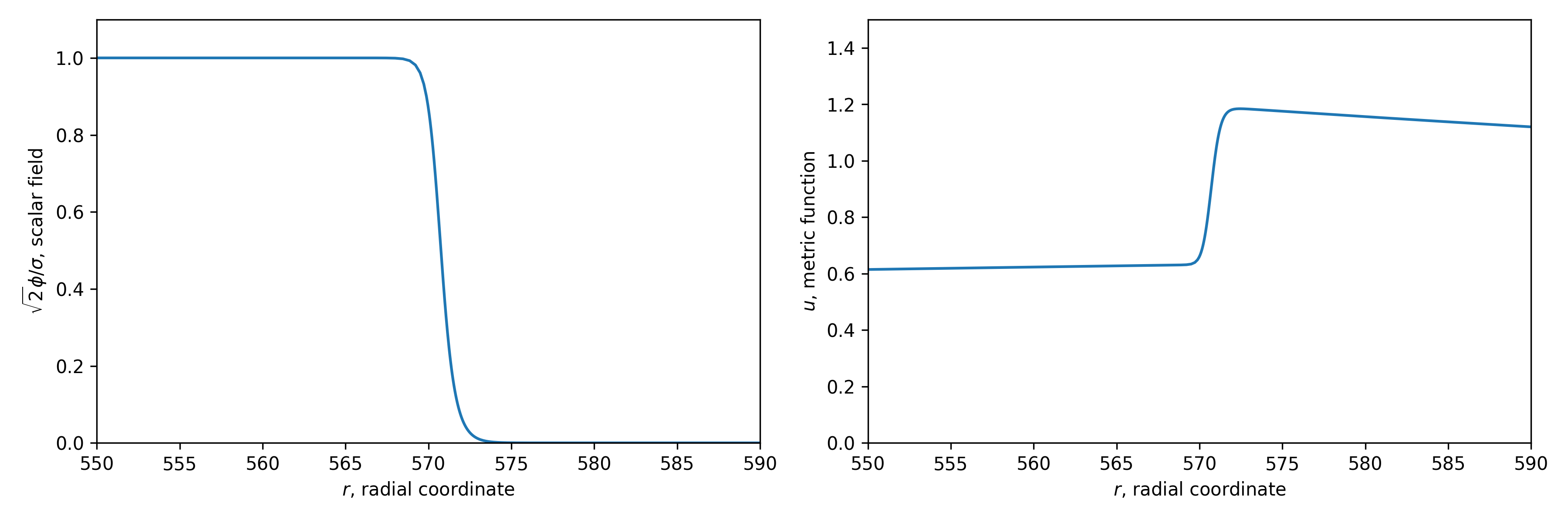}
	\end{center}
	\caption{Zoom in into the stiffest part of the scalar field profile for a very stiff solitonic star solution (with the potential parameter $\sigma_0 = 0.05$)}
	\label{Fig:spherical_star_zoom_in}
\end{figure}

We have computed the stable sections of the fundamental branches of spherically symmetric boson stars for the potential parameters $\sigma_0 = \{0.05, 0.1, 0.2, 0.4\}$. We present the $\phi_c$ vs $\Omega$, $M$ vs $\Omega$, $Q/M^2$ vs $\Omega$ and compactness $\mathcal{C} = M/R$ vs $\Omega$ diagrams on the Figure~\ref{Fig:spherical_star_tracing}. The $\phi_c$ stands for the absolute value of the central value of the scalar field, which also turns out to be a global maximum of the absolute value of the scalar field. 
The results largely reproduce the analytical approximation results \citep{Friedberg:1986tq} and the numerical explorations in \citep{Collodel:2022jly}, however we could go well past the stiffness explored in \citep{Collodel:2022jly}. An example of a star with the potential parameter $\sigma_0 = 0.05$ near the mass extremum is given on the Figure~\ref{Fig:spherical_star_example}, the zoom in on the transition region is given on the Figure~\ref{Fig:spherical_star_zoom_in}.
We see that the multidomain Chebyshev collocation method is extremely effective in exploring stiff solutions. The particular choice of $\sigma_0 = 0.05$ for the stiffest potential has no significance, the method allows us to go much stiffer families, however, even the families we present are very well described by the analytic approximation \citep{Friedberg:1986tq}.

The extreme success of our implementation of the multidomain Chebyshev collocation method can be partially attributed to a very simple geometry of the solution. Indeed, the sharp transition of the spherically symmetric boson star solution happens in the region that can be put entirely into a single interval. As we have noted before, the domain decomposition for the rotating solitonic stars is much more complicated (see Figure~\ref{Fig:stiff_solution}).

\section{Conclusions}
In this paper we present a new method for constructing boson star solutions with a stiff almost discontinuous profiles in general relativity. The method covers the computational domain with a set of rectangular blocks with completely separate collocation grids set on each one and simultaneously solves for the equations inside the bulk of the domains together with appropriate inter-domain matching conditions. We set the equations in spherical coordinates and extend the solution in parity odd or even way across the origin and the axis of symmetry to avoid setting boundary conditions near coordinate singularities of spherical coordinates.

We apply the method to axisymmetric rotating solitonic boson stars and spherically symmetric non-rotating solitonic boson stars. We demonstrate self-consistency of the method by studying $L_\infty$ convergence of independent residuals and showing the convergence to be exponential. The broad investigation of the solitonic boson stars will be presented in the follow-up paper \citep{Sukhov:bstars2}.

In the non-rotating solitonic boson star case we numerically explore the solitonic thin-shell limit that was previously studied in \citep{Friedberg:1986tq} with an analytical approximation, our numerical results confirm the results of \citep{Friedberg:1986tq} and show that our method can treat arbitrarily sharp star profiles with little effort. We also present a few useful techniques for numerical exploration of boson star families.

Our method can readily be applied to other systems of non-linear coupled elliptic equations in arbitrary number of dimensions provided that the solution is sufficiently smooth and differentiable, although in terms of computational resource allocation it is practical in 1, 2 and 3 dimensions. Our software implementation of the method makes it extremely easy to apply, one simply needs to plug in the equations with appropriate boundary conditions and use a good initial guess for relaxation.

\section{Acknowledgments}
We are hugely indebted to Frans Pretorius for his patient guidance throughout this project. We also thank Justin Ripley, Alex Pandya, Alexey Milekhin and Oleksandr Stashko for useful discussions.

\section{Appendix}
\subsection{Rotating Boson Star equations}
The substitution of the ansatz \eqref{Description:KEH_ansatz} directly into the Einstein equations \eqref{Description:Einstein_equations} leads to the system of four equations and two independent residuals. The equations can be rearranged into the following system
\begin{subequations}
	\begin{align}
		&\omega_{,rr} + \left(\frac{4}{r} + \gamma_{,r} - 2\rho_{,r}\right) \omega_{,r} + \frac{1}{r^2}\left[\omega_{,\theta\theta}+\left(3\frac{\cos\theta}{\sin\theta}+\gamma_{,\theta}-2\rho_{,\theta}\right)\omega_{,\theta}\right] \\\nonumber
		&- 2\kappa e^{\frac{1}{2}(\gamma + \rho) + 2\sigma} j^\varphi = 0,\\\nonumber
		&\gamma_{,rr}+\left(\frac{3}{r}+\gamma_{,r}\right)\gamma_{,r}+\frac{1}{r^2}\left[\gamma_{,\theta\theta}+\left(2\frac{\cos\theta}{\sin\theta}+\gamma_{,\theta}\right)\gamma_{,\theta}\right]\\\nonumber
		&- \kappa e^{2\sigma}(2\rho_e - \hat{S}^r_{\;r} - \hat{S}^\theta_{\;\theta} - 2\hat{S}^\varphi_{\;\varphi}) = 0,\\
		&\rho_{,rr} + \left(\frac{2}{r}+\gamma_{,r}\right)\rho_{,r}+\frac{1}{r^2}\left[\rho_{,\theta\theta}+\left(\frac{\cos\theta}{\sin\theta}+\gamma_{,\theta}\right)\rho_{,\theta}\right]-\frac{1}{r}\left(\gamma_{,r}+\frac{1}{r}\frac{\cos\theta}{\sin\theta}\gamma_{,\theta}\right)\\\nonumber
		&- e^{-2\rho}(r^2 \omega_{,r}^2 + \omega_{,\theta}^2)\sin^2\theta - \kappa e^{2\sigma}(2\rho_e - \hat{S}^r_{\;r} - \hat{S}^\theta_{\;\theta}) = 0,\\
		&\sigma_{,rr}+\frac{1}{r}\sigma_{,r} + \frac{1}{r^2}\sigma_{,\theta\theta} - \frac{1}{4}\left(\frac{2}{r}+\gamma_{,r}\right)\gamma_{,r}-\frac{1}{4 r^2}\left(2\frac{\cos\theta}{\sin\theta}+\gamma_{,\theta}\right)\gamma_{,\theta}\\\nonumber
		&-\frac{1}{4}\left(\frac{2}{r}-\rho_{,r}\right)\rho_{,r}-\frac{1}{4r^2}\left(2\frac{\cos\theta}{\sin\theta}-\rho_{,\theta}\right)\rho_{,\theta} -  \frac{1}{4}e^{-2\rho}(r^2 \omega_{,r}^2 + \omega_{,\theta}^2)\sin^2\theta\\\nonumber
		&+ \kappa e^{2\sigma} (\rho_e - \hat{S}^\varphi_{\;\varphi}) = 0,
	\end{align}
\end{subequations}
where
\begin{equation}
	\rho_e = T_{ab} n^a n^b,\qquad j^\phi = - n^a T_a^{\;\phi},\qquad \hat{S}_{ik} = T_{ik} - \frac{1}{2}\gamma_{ik}\left(T_{kl}\gamma^{kl} - \rho_e\right),
\end{equation}
and $n^a$ is the unit normal vector to $t = const$ family of hypersurfaces.

The two remaining residuals are
\begin{subequations}
	\label{Model:residual_equations}
	\begin{align}
		\label{Model:first_residual}
		&\gamma_{,r\theta} + \frac{1}{2}\gamma_{,r}\gamma_{,\theta} + \left(\frac{1}{2}\frac{\cos\theta}{\sin\theta} - \sigma_{,\theta}\right)\gamma_{,r} - \left(\frac{1}{2r}+\sigma_{,r}\right)\gamma_{,\theta}-\frac{1}{2}\frac{\cos\theta}{\sin\theta}\rho_{,r}+\frac{1}{2}\rho_{,r}\rho_{,\theta}\\\nonumber
		&-\frac{1}{2r}\rho_{,\theta}  - \frac{\cos\theta}{\sin\theta}\sigma_{,r} - \frac{1}{r}\sigma_{,\theta} - \frac{1}{2}e^{-2\rho}\omega_{,r}\omega_{,\theta}\, r^2 \sin^2\theta + \kappa e^{2\sigma} \hat{S}^r_{\,\theta} = 0,\\
		\label{Model:second_residual}
		&\gamma_{,rr}+\frac{1}{2}\left(\frac{2}{r}+\gamma_{,r}\right)\gamma_{,r}+\sigma_{,rr}-\gamma_{,r}\sigma_{,r}+\frac{1}{r^2}\left(\sigma_{,\theta\theta}+\left(\frac{\cos\theta}{\sin\theta}+\gamma_{,\theta}\right)\sigma_{,\theta}\right)\\\nonumber
		&-\frac{1}{2}\left(\frac{2}{r}-\rho_{,r}\right)\rho_{,r}-\frac{1}{2}e^{-2\rho}\omega_{,r}^2 r^2 \sin^2\theta + \kappa e^{2\sigma} \hat{S}^r_{\; r} = 0.
	\end{align}
\end{subequations}

We now consider the components of stress-energy tensor \eqref{Description:stress-energy tensor} for the scalar field ansatz \eqref{Description:phi_ansatz}. We get the following energy density $\rho_e$
\begin{equation}
	\rho_e = e^{-\gamma}\left[e^{-\rho}(\Omega + m\, \omega)^2 + e^\rho\frac{m^2}{r^2\sin^2\theta} \right]\phi^2 + e^{-2\sigma}\left(\phi_{,r}^2 + \frac{1}{r^2}\phi_{,\theta}^2\right) + V.
\end{equation}
The momentum current has only one surviving component $j^\phi$
\begin{equation}
	j^\phi = \frac{2}{r^2 \sin^2\theta} e^{\frac{1}{2}\rho - \frac{3}{2}\gamma} m (\Omega + m\,\omega) \phi^2.
\end{equation}
The non-zero components of stress-energy tensor $\hat{S}_{ik}$ are
\begin{subequations}
	\label{Model:scalar_stress_energy}
	\begin{align}
		&\hat{S}^r_{\; r} = 2 e^{-2\sigma} \phi_{,r}^2 + V,\\
		&\hat{S}^r_{\; \theta} = 2 e^{-2\sigma} \phi_{,r}\phi_{,\theta},\\
		&\hat{S}^\theta_{\; \theta} = \frac{2}{r^2} e^{-2\sigma} \phi_{,\theta}^2 + V,\\
		&\hat{S}^\varphi_{\; \varphi} = \frac{2 m^2}{r^2 \sin^2\theta} e^{-\gamma+ \rho}\phi^2 + V.
	\end{align}
\end{subequations}

The Klein-Gordon equation for the scalar field \eqref{Description:KG_equation} on the metric \eqref{Description:KEH_ansatz} backgound becomes
\begin{equation}
	\label{Model:scalar_field_equation}
	\begin{split}
		&\phi_{,rr} + \left(\frac{2}{r}+\gamma_{,r}\right)\phi_{,r}+\frac{1}{r^2}\left(\phi_{,\theta\theta} + \left(\frac{\cos\theta}{\sin\theta}+\gamma_{,\theta}\right)\phi_{,\theta}\right)\\
		&+e^{2\sigma}\bigg[e^{-\gamma}\left(e^{-\rho}(\Omega + m\, \omega)^2 - e^\rho \frac{m^2}{r^2\sin^2\theta}\right)-V'\bigg]\phi = 0.
	\end{split}
\end{equation}

\subsection{Static Boson Star equations}
The substitution of the spherically symmetric ansatz \eqref{Description:Schwarzschild metric} directly into the Einstein equations \eqref{Description:Einstein_equations} leads to the system of two equations
\begin{subequations}
	\label{1Dstar:gravity_EOM}
	\begin{align}
		\label{1Dstar:gravity_u_equation}
		\partial_r u = \frac{1}{r}\left(1 - e^u (1 - \kappa r^2 \rho)\right),\\
		\partial_r v = \frac{1}{r}\left(e^u (\kappa r^2 p + 1) - 1\right),
	\end{align}
\end{subequations}
and a single independent residual
\begin{equation}
	\partial_r^2 v + \frac{1}{2}(\partial_r v - \partial_r u)\left(\frac{2}{r} + \partial_r v\right) = 2 \kappa e^{u} p_\theta,
\end{equation}
where the energy density, the pressure and the angular pressure are given by
\begin{equation}
	\begin{aligned}
		&\rho = \Omega^2 e^{-v} \phi_0^2 + e^{-u} (\partial_r\phi_0)^2 + V(\phi_0^2),\\
		&p = \Omega^2 e^{-v} \phi_0^2 + e^{-u} (\partial_r\phi_0)^2 - V(\phi_0^2),\\
		&p_\theta = \Omega^2 e^{-v} \phi_0^2 + e^{-u} (\partial_r\phi_0)^2 - V(\phi_0^2).
	\end{aligned}
\end{equation}
The Klein-Gordon equation for the scalar field \eqref{Description:KG_equation} on the Schwarzschild-like metric \eqref{Description:Schwarzschild metric} backgound becomes
\begin{equation}
	\label{1Dstar:scalar_field_equation}
	\partial_r^2\phi_0 + \left(\frac{2}{r} + \frac{\partial_r v - \partial_r u}{2}\right)\partial_r \phi_0 = e^u (U_0 - \Omega^2 e^{-v})\phi_0,\qquad U_0 = \frac{d V}{d|\phi|^2}
\end{equation}

\bibliographystyle{plain}
\bibliography{biblography}

\end{document}